\documentclass[prb,twocolumn,showpacs,preprintnumbers,amsmath,amssymb]{revtex4}
\usepackage{bm}
\usepackage{graphics}
\usepackage{dcolumn}
\usepackage{latexsym}

\begin{document}

\title{Phase transitions in random magnetic bilayer}

\author{P. N. Timonin}
\email{timonin@aaanet.ru}
\author{ V. B. Shirokov }
\affiliation{Physics Research Institute at Rostov State University
344090, Rostov - on - Don, Russia}

\date{\today}

\begin{abstract}
The influence of random interlayer exchange on the phase states of the simplest magnetic heterostructure consisting of two ferromagnetic Ising layers with large interaction radius is studied. It is shown that such system can exist in three magnetic phases: ferromagnetic, antiferromagnetic and ferrimagnetic. The possible phase diagrams and temperature dependencies of thermodynamic parameters are described. The regions of existence of the magnetic phases in external magnetic field are determined at zero temperature.
\end{abstract}


\maketitle

Thin films of layered magnets and artificial heterostructures composed of alternating magnetic and nonmagnetic layers can have a variety of stable magnetic states and switching between them can be achieved by different regimes of magnetic field variation \cite{1,2,3}. This opens vast possibilities for numerous technical applications of such structures \cite{4,5,6} and makes theoretical description of their magnetic states and properties of these states very actual. In particular, such description for finite number of layers can be achieved in some variants of mean-field approximations \cite{6,7}  or using numerical methods \cite{8,9}, yet the study of ideal homogeneous systems can be insufficient for the description of experimental data. Indeed, impurities and defects in interlayer space modify generally the exchange constants, which could essentially transform magnetic state of layered structure. Fluctuations of nonmagnetic spacer thickness also give rise to the similar effect of local interlayer exchange modification \cite{10,11}.

Here we consider the influence of random interlayer exchange on the phase states of the simplest magnetic heterostructure consisting of two ferromagnetic Ising layers. It can be realized as thin two-layer crystal film having controlled concentration of impurities in the interlayer space or as two monoatomic layers deposited on the surfaces of nonmagnetic spacer with random thickness fluctuations. We consider the case when radius of intralayer ferromagnetic interaction is much larger than lattice parameter. Then the model can be treated in the mean-field approximation (except of narrow vicinity of transition point), i. e. the intralayer interaction radius can be put infinite. In such approximation it is quite easy to obtain the expression for averaged thermodynamic potential depending on the layers' magnetizations, establish possible magnetic phases of such system and describe its thermodynamic properties in these phases.

\section{Thermodynamic potential of random magnetic bilayer}

The Hamiltonian of the model has the form

\begin{eqnarray}
\begin{array}{l}
{\cal H} =  - \frac{J}{{2N}}\sum\limits_{i = 1}^2 {\left( {\sum\limits_{n = 1}^N {S_{n,i} } } \right)^2 }  -  \sum\limits_{n = 1}^N {\tilde J_n S_{n,1} S_{n,2} } \\
 - H\sum\limits_{n = 1}^N {\sum\limits_{i = 1}^2 {S_{n,i} } } \label{eq:1}
 \end{array}
\end{eqnarray}

Here $S_{n,i}  =  \pm 1$ are Ising spins, index $n$ numerates lattice cites in layers from 1 to $N$, $i = 1, 2$ is the layer number, $J > 0$  is the constant of intralayer ferromagnetic exchange, $\tilde J_n $ is random interlayer exchange, $H$ is external magnetic field. We assume that all $\tilde J_n $ have the common distribution function

\begin{equation}
W(\tilde J) = \left( {1 - p} \right)\delta \left( {\tilde J - J_ +  } \right) + p\delta \left( {\tilde J - J_ -  } \right). \label{eq:2} \\[6pt]
\end{equation}

It describes the presence of impurities or defects of one sort with concentration $p$, which change the interlayer interaction constant on sites from $J_ + $ in pure system to $J_ -  $. We can put $J_ +   > J_ -  $ without loss of generality.

According to Ref. [\onlinecite{12}] the averaged over $\tilde J_n $ inequilibrium potential (per one site in the layer) corresponding to Eq. (\ref{eq:1}) is

\begin{equation}
\begin{array}{l}
F\left( {\bf m} \right) =  J{\bf m}^2 /2  \\[6pt]
\qquad   -T\left\langle {\ln \left\{ {Tr\exp \left[ {\tilde J S_1 S_2  + \sum\limits_{i = 1}^2 {\left( {H + Jm_i } \right)S_i } } \right]/T} \right\}} \right\rangle. \label{eq:3} \\[6pt]
\end{array}
\end{equation}

Here $m_1$ and $m_2$ are the layers' magnetizations, $T$ - temperature, $Tr$ designates the sum over spin configurations and the angular brackets mean the average over random $\tilde J$.
Minimizing the potential in Eq. (\ref{eq:3}) over $m_1$ and $m_2$ one can find their equilibrium values as functions of $T$, $p$, $H$ and determine all thermodynamic propeties of the model. Summing over spins in Eq. (\ref{eq:3})  and using Eq. (\ref{eq:2})  we get the explicit form of $F(\bm m)$,

\begin{widetext}
\begin{equation}
\begin{array}{l}
 F\left( {\bf m} \right) =  - T\ln 2 + m^2  + l^2 -Tp\ln \left[ {e^{J_ -  /T} \cosh\left( {\frac{{2m + 2H}}{T}} \right) + e^{ - J_ -  /T} \cosh\left( {\frac{{2l}}{T}} \right)} \right] \\[6pt]
\qquad \qquad - T\left( {1 - p} \right)\ln \left[ {e^{J_ +  /T} \cosh\left( {\frac{{2m + 2H}}{T}} \right) + e^{ - J_ +  /T} \cosh\left( {\frac{{2l}}{T}} \right)} \right] \label{eq:4} \\[6pt]
\end{array}
\end{equation}
\end{widetext}

Here $m = \left( {m_1  + m_2 } \right)/2$ is full magnetization and $l = \left( {m_1  - m_2 } \right)/2
$ is antiferromagnetic order parameter. In Eq. (\ref{eq:4}) the unit system is used in which $ J = 1$.

\begin{figure*}
\includegraphics{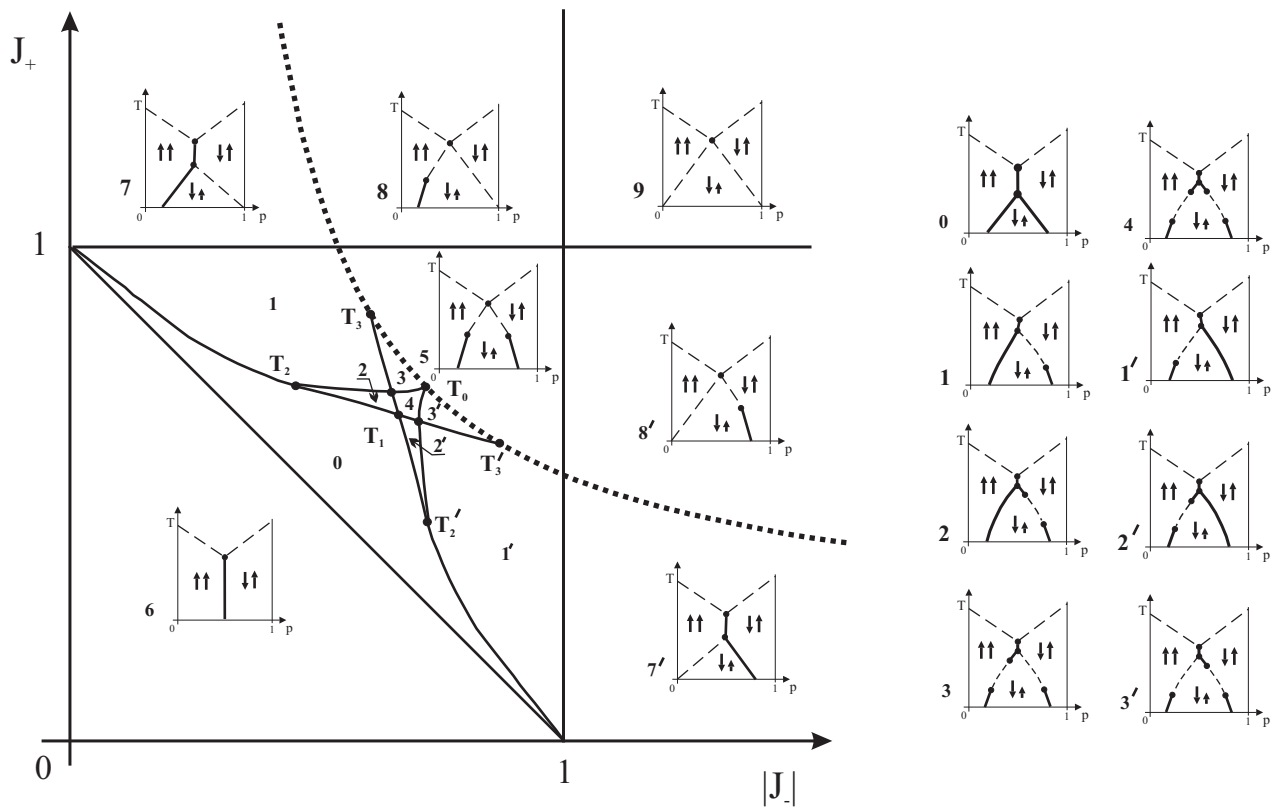}
\caption{\label{Fig.1} The sections of the excange parameter space with different types of (p, T)
 phase diagrams shown in the inserts. Solid and dashed lines designate first- and second-order
transitons correspondingly. The points' coordinates are: $T_0-(0.658, 0.658)$, $T_1-(0.619, 0.619)$, $T_2-(0.418, 0.690)$, $T'_2-(0.690, 0.418)$, $T_3-(0.518, 0.868)$, $T'_3-(0.868, 0.518)$.}
\end{figure*}

\section{Phase transitions in zero field.}

The potential in Eq. (\ref{eq:4}) describes generally the competition between ferromagnetic and antiferromagnetic ordering of the layers. Equating to zero the coefficients at $m^2$ and $l^2$ in its expansion we find the equations for the temperatures of the ferromagnetic transition ($T_+$) and the antiferromagnetic one ($T_-$) correspondingly

\begin{equation}
\begin{array}{l}
T_ +   = 1 + \left( {1 - p} \right)\tanh\left( {\frac{{J_ +  }}{{T_ +  }}} \right) + p\tanh\left( {\frac{{J_ -  }}{{T_ +  }}} \right) , \\[6pt]
T_ -   = 1 - \left( {1 - p} \right)\tanh\left( {\frac{{J_ +  }}{{T_ -  }}} \right) - p\tanh\left( {\frac{{J_ -  }}{{T_ -  }}} \right) .  \label{eq:5} \\[6pt]
\end{array}
\end{equation}

The second order phase transition from paramagnetic phase $m=0$, $l=0$ takes place into the phase which have the higher transition temperature. Hence we have transition into the ferromagnetic (F) phase for $J_ \pm   > 0$ and transition into the antiferromagnetic (A) phase  for $J_ \pm   < 0$  at all $p$. In both these cases having trivial thermodynamics with just one equilibrium phase, the disorder can induce the appearance of other maghetic phases only as metastable ones. This can only cause some pecularities in the kinetics of the system.

Naturally, the real competition of two types of equilibrium ordering is present when $J_ +   > 0,{\rm    }J_ -   < 0$. Then there is the critical concentration $p_c$ below which ferromagnetic transition takes place while above it antiferromagnetic transition appears. From Eqs. (\ref{eq:5}) we have

\begin{equation*}
\begin{array}{l}
p_c  = \frac{{\tanh (J_ +)  }}{{\tanh (J_ +)   + \tanh\left| {J_ -  } \right|}}, \\[6pt]
T_ +  \left( {p_c } \right) = T_ -  \left( {p_c } \right) = 1. \\[6pt]
\end{array}
\end{equation*}

Below we focus mainly on this case of competing random interactions.

Numerical analysis of the $F(\bf m)$ minima shows that on the ($p$, $T$) -  phase  diagram the point ($p_c$, 1) is multiphase one, in its vicinity three or four magnetic phases can exist. The fourth phase is ferrimagnetic (FA) one where both $m$ and $l$ are nonzero, so layers have different magnetization modules. The appearance of this phase is the main qualitative effect of random interlayer exchange, it does not appear without such randomness.

To elucidate the macroscopic mechanism underlying the appearance of such states in the presence of just small fraction of negative interlayer bonds one can consider the possible ground states describing the properties of the system at $T=0$. The simple analytical calculations can show how spins may arrange to provide the states with $m_1 \neq m_2$ minimizing the energy in Eq. (\ref{eq:1}). Indeed, from  Eq. (\ref{eq:1}) we have for the energy density in some specific spin and interlayer bonds configuration
\[
E =  - \left( {m_1^2  + m_2^2 } \right)/2 + \left| {J_ -  } \right|\Sigma _ -   - J_ +  \Sigma _ +
\]
Here
\[
\Sigma _ -   = N^{ - 1} \sum\limits_{n = 1}^{pN} {S_{n,1} S_{n,2} } ,\text{           }\Sigma _ +   = N^{ - 1} \sum\limits_{n = 1}^{\left( {1 - p} \right)N} {S_{n,1} S_{n,2} }
\]
with the sums running over the sites with negative and positive interlayer bonds in $\Sigma _ - $ and $\Sigma _ + $ correspondingly.
When negative bonds $J_-$ are absent ($p=0$) the lowest energy has ferromagnetic spin state with all spins, say, up. Introducing negative bonds with some small concentration could make some spins coupled by such bonds to point down in the ground state. Consider the state with $N_1$ reversed spins in first layer and $N_2$ reversed spins in the second one belonging to the different pairs. Then we have
\[
\begin{gathered}
  \Sigma _ -   = p - 2n_1  - 2n_2 ,\text{   }\Sigma _ +   = 1 - p, \hfill \\
  m_i  = 1 - 2n_i,\text{    } n_i  = N_i /N, \text{  }i = 1,2, \hfill \\
\end{gathered}
\]
with $n_i  > 0$ and $n_1  + n_2  < p$.

Substituting these expressions in $E$ one can see that the resulting $E(n_1, n_2)$ has no minima inside the region where $n_i$ are defined. So the minima are present at the boundaries of this region. It is easy to verify that there could be two absolute minima, ferromagnetic one ($n_1 = n_2 =0$) with
\[
E(0,0) =  - 1 + p\left| {J_ -  } \right| - \left( {1 - p} \right)J_ +
\]
and ferrimagnetic minimum at $n_1 = p, n_2 =0$ or $n_1 = 0, n_2 = p$ with
\[
E(p,0) =  - \left[ {\left( {1 - p} \right)^2  + p^2 } \right] - p\left| {J_ -  } \right| - \left( {1 - p} \right)J_ + .
\]
Apparently, the ground state is the ferrimagnetic one ($m_1 = 1- 2p, m_2 = 1$ or $m_1 = 1, m_2 = 1- 2p$) if $\left| {J_ -  } \right| > 1 - p$.

The global ($p$, $T$) - phase diagrams presented in Fig. (\ref{Fig.1}) for the whole exchange parameter space show that the FA phase can adjoin the paramagnetic one only at the point ($p_c$, 1). Generally the transition into it is possible only from F and  A phases and it can be either of second - or first - order type.

The dotted curve in Fig. (\ref{Fig.1}) divides the exchange parameter space in two regions with phase diagrams having four-phase point and only three-phase point at ($p_c$, 1). This curve is defined by the equation $ 3\tanh (J_ +)  \tanh\left| {J_ -  } \right| = 1$ and below it the stability regions of F and A phases always overlap resulting in the first-order transition between them.
At $J_+  > 1$ F-FA transition is always of the second-order type and the same is true for A-FA transition for $\left| {J_ -  } \right| > 1$. When $ J_+ +\left| {J_ -  } \right| < 1$ the FA phase is always metastable.
The regions 0-4 and 1'-3' differ by the number of tricritcal points on the boundaries of FA phase at which the second - and first - order transitions meet. The lines between these regions are given by cumbersome parametric equations, which we do not present here.

\begin{figure*}
\includegraphics{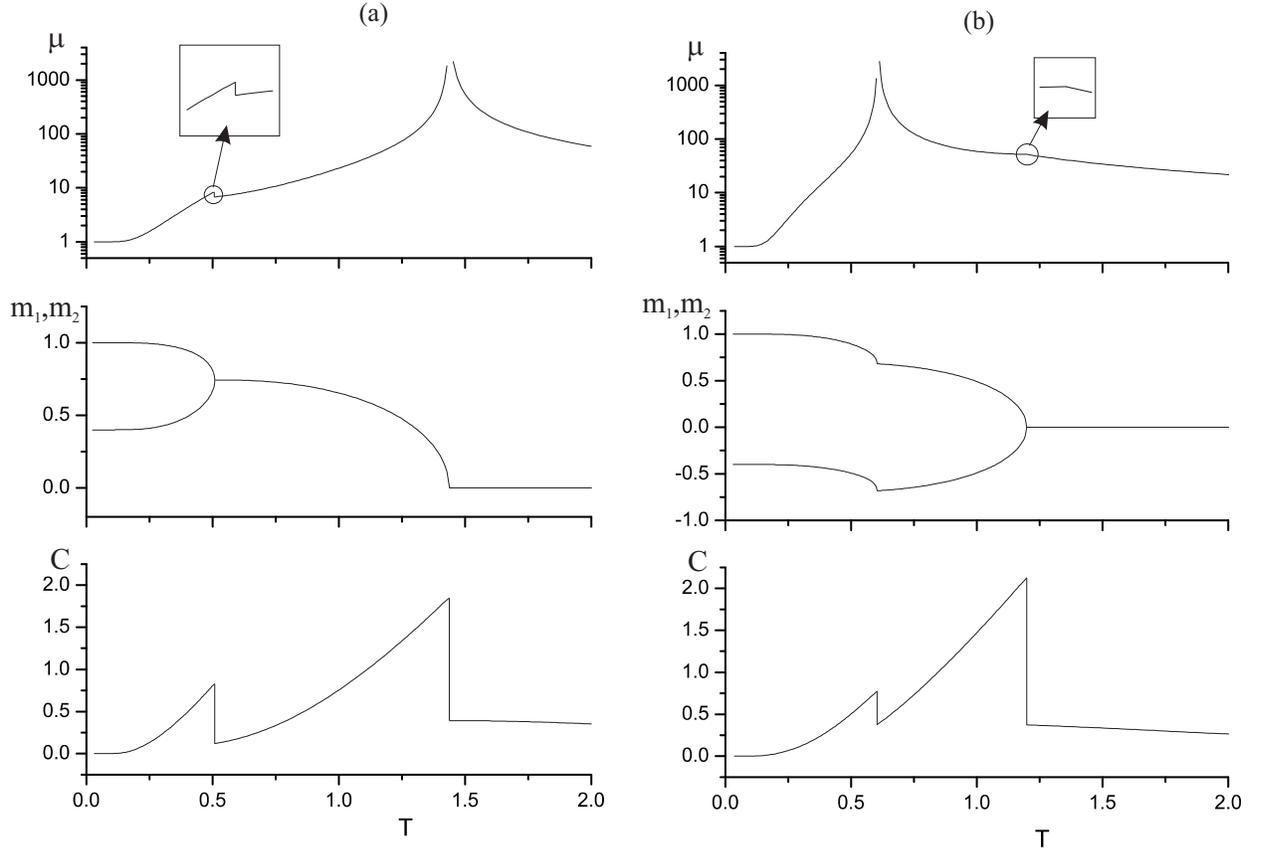}
\caption{\label{Fig.2} Temperature dependencies of magnetic permeability, layers' magnetizations
and heat capacity in the region 9 of Fig. (\ref{Fig.2}), (a)- $p=0.3$, (b)- $p=0.7$ }
\end{figure*}

Fig. (\ref{Fig.2}) shows the typical temperature dependencies of the magnetic permeability, layers' magnetizations and heat capacity. They exhibit the characteristic mean-field anomalies at the second-order phase transitions. Note that magnetic permeability is not critical at the transitions into A phase and from F to FA phase, so it has a break in the first case and a jump in the second one.

\section{Magnetic phases in external magnetic field at T=0}

The study of phase states in magnetic field in the system considered is quite simple at $T=0$. Here we can get analytical expressions for their stability regions. Indeed at $T=0$ we have from Eq. (\ref{eq:3}) (putting again $J=1$)

\begin{widetext}
\begin{equation}
\begin{array}{l}
 F\left( {\bf m} \right) = {\bf m}^2 /2 - \left\langle {\mathop {\min }\limits_{S_1 ,S_2 } \left[ {\tilde JS_1 S_2  + \sum\limits_{i = 1}^2 {\left( {H + m_i } \right)S_i } } \right]} \right\rangle   \\[12pt]
\qquad  \quad  = m^2  + l^2  - \left\langle {\left( {\tilde J + 2\left| {m + H} \right|} \right)\vartheta \left( {\tilde J + \left| {m + H} \right| - \left| l \right|} \right)} \right\rangle   +
 \left\langle {\left( {\tilde J - 2\left| l \right|} \right)\vartheta \left( {\left| l \right| - \tilde J - \left| {m + H} \right|} \right)} \right\rangle \label{eq:6} \\[6pt]
\end{array}
\end{equation}
\end{widetext}

Here $\vartheta \left( x \right)$  is Haeviside's step function.
The form of the potential in Eq. (\ref{eq:6}) implies that all its extrema are minima, so all solutions to eqations of state correspond to stable phases. Their stability regions and the equilibrium potential values are:

1. F phases,  $m = \pm 1$, $l = 0$,
\begin{equation*}
\begin{array}{l}
Hsign(m) > \max (H_f , - 1), \\[6pt]
H_f  \equiv  - 1 - J_ -  ,\\[6pt]
F_1  =  - 1 - pJ_ -  - (1 - p)J_ +  - 2{\rm  }Hsign(m) .\\[6pt]
\end{array}
\end{equation*}

2.  A phases,  $m = 0$, $l = \pm 1$,
\begin{equation*}
\begin{array}{l}
\left| H \right| < H_a  \equiv 1 - J_ + ,\\[6pt]
F_2  =  - 1 + pJ_ -  + (1 - p)J_ +\\[6pt]
\end{array}
\end{equation*}

3. FA phases $m = \pm (1 - p)$, $l = p$,
\begin{equation*}
\begin{array}{l}
H×sign(m) + \left| m \right| > 0, \\[6pt]
H_a  < Hsign(m) + 2{\rm  }\left| m \right| < H_f  + 2, \\[6pt]
F_3 =  - p^2  - (1 - p)^2  + pJ_ -  - (1 - p)J_ +  - 2mH \\[6pt]
\end{array}
\end{equation*}

\begin{figure}
\includegraphics{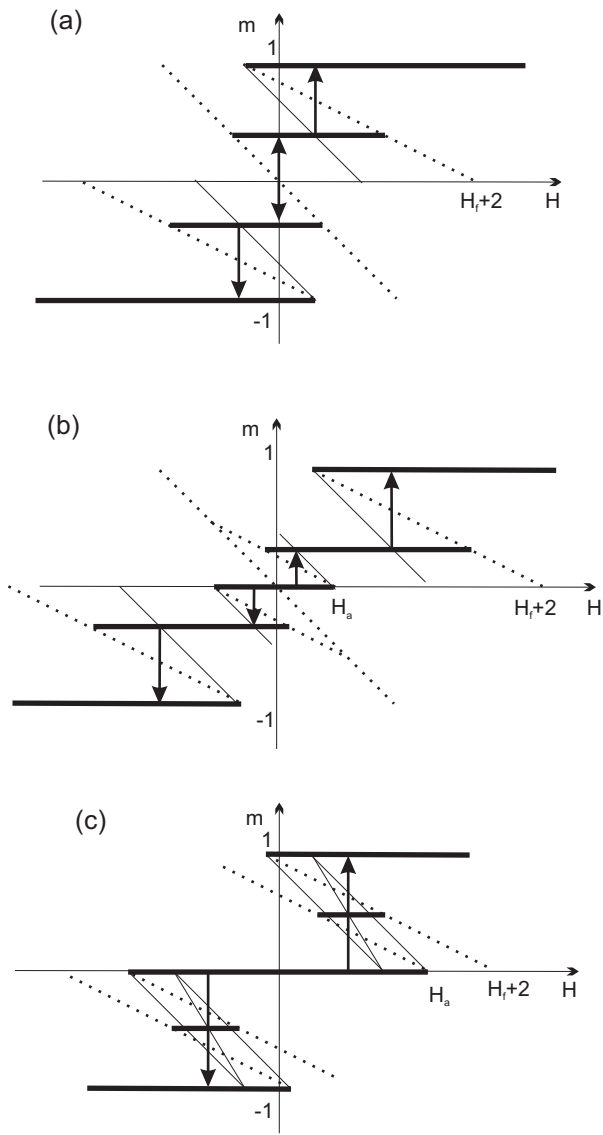}
\caption{\label{Fig.3} Magnetic states in external field at $T=0$ for $H_f >-1$. (a) $H_a < 0$, (b) $0 < H_a < 1$, (c) $1 < H_a$. The dotted lines show the boundaries of the existence of FA states at varying $p$. Thin solid lines are used to determine the first-order transition points shown by the arrows. The tangents of the inclined lines are -1 and -1/2 except the line on Fig. 3(c) intersection of which with FA magnetization gives the A-F transition point.}
\end{figure}

\begin{figure}
\includegraphics{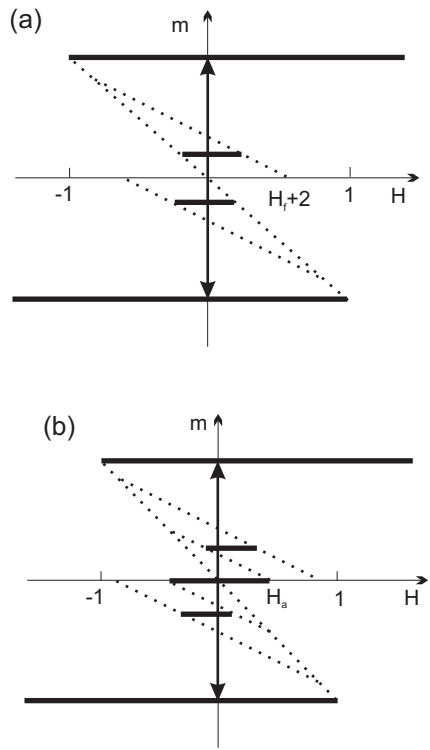}
\caption{\label{Fig.4} The same that in Fig. (\ref{Fig.3}) for $H_f <-1$. (a) $ H_a < 0$, $-2 < H_f  < -1$,  (b) $0 < H_a < 1$}
\end{figure}

The stability regions of these phases and their magnetizations $m$ are shown in Fig. (\ref{Fig.3}) for $H_f >-1$ and in Fig. (\ref{Fig.4})  for $H_f <-1$. Here the FA magnetization is shown for only one $p$ value, while moving it inside the band bounded by dotted lines one can obtain its position for another $p$.
Note that in Fig. (\ref{Fig.4}) the cases with $J_\pm >0$ are shown, which have the equilibrium F phase at all $H$, while A and FA phases appeare only as metastable ones.

The first-order transitions can take place between coexisting phases at the fields in which their potentials become equal. Let us define the fields

\begin{equation*}
\begin{array}{l}
H_1 =  H_a -1 +p, \\[6pt]
H_2 = H_f +p, \\[6pt]
H_3 = (1-p)H_1 +pH_2. \\[6pt]
\end{array}
\end{equation*}

When $H_1 < 0 < H_2$ there are FA-FA transition at $H=0$ and FA-F transitions at $H = \pm H_2$ while for $0 < H_1 < H_2$ we have A-FA and FA-F transition at $H = \pm H_1$ and $H = \pm H_2$ correspondingly. If $H_3 < H_1$ just A-F transitions take place at $\pm H_3$. In other cases there are only F-F transition at $H=0$, see Fig. (\ref{Fig.4}). The simple graphical procedure for obtaining these transition points is shown in Fig. (\ref{Fig.3}).

The graphs in Figs. (\ref{Fig.3}), (\ref{Fig.4}) give also a notion of the possible forms of hysteresis loops in slow varying periodic fields. In this case the transitions between magnetic states appear at the points where phases seize to exist since the period of field oscillations is usually much less than the decay time of metastable state. This implies also that hysteresis loops' forms could depend on the initial state as well as the amplitude of the applied field. So for each graph in Figs.(\ref{Fig.3}), (\ref{Fig.4}) we could have several loops. Yet the present static data are insufficient to predict their exact forms in the cases when several stable states exist at a given field. Then one should consider kinetics as it is not generally evident to which state the system drops. For example, the exisence of small loops at low field amplitudes inside the large ferromagnetic one in Fig. (\ref{Fig.4}) can only be proved in the kinetic framework. Note also that the coercive field defining the boundary of ferromagnetic state is never greater than the one-layer coercive field being 1 in the present system of units.

It is easy to envisage the evolution of the graphs in Figs. (\ref{Fig.3}), (\ref{Fig.4}) with the temperature rising. Thus the growth of permeability, see  Fig. (\ref{Fig.2}), means that the lines depicting magnetizations become inclined while the field intervals where they exist become smaller. Finally the FA phases vanish at some temperature and above it only usual ferromagnetic and antiferromagnetic (double) loops would exist.

\section{Discussion}
The above theory can also be valid when instead of two monoatomic layers there are two slabs each having $k$ such layers. Suppose the layers in the slabs are tightly bound by the nearest-neighbor ferromagnetic exchange($J'$) which is much stronger than intralayer ($J$) and random spacer ($\tilde J$) ones, $J'\gg (kJ, \tilde J)$. Then  all layers in the slabs would have parallel magnetizations at $T \ll J'$ so the present results hold after the substitution $J \rightarrow kJ$. While such slabs are ubiquitous experimental objects, only the different ones were usually studied in bilayers, see Refs. [\onlinecite{1,2,3,4,5,9,13,14}]. Evidently this is due to expected triviality of results (just one magnetic state) for the identical slabs without randomness (yet see Ref. [\onlinecite{6}]).
The present study show that random competing exchange can essentially complicate the variety of phase states in the simplest layered heterostructure. In case of different layers (slabs) the phase diagrams would have up to four ferrimagnetic phases as preliminary study shows. The variety of magnetic states in random multilayer structures will be still more complex, and one may expect a fast phase alternation under small changes of external parameters. This could be seen in experiment as the region of spin-glass state \cite{15}.

Yet additional magnetic states in random heterostructures could expand the possibilities of their technical applications. Their study could also be fruitful for the general theory of disordered magnets. Thus the simplest system considered here allows one to conclude that the influence of competing random exchange does not come down to the changes of critical indexes. It can also cause the appearance of new magnetic phases and transitions between them under temperature as well as disorder variations. It is natural to expect the existence of similar effects in the most studied objects with random competing exchange such as spin-glasses with that distinction that they would have enormous multitude of phases and phase transitions.

\begin{acknowledgments}
This work was made under support from Russian Foundation for Basic Researches, grants 04-02-16228, 06-02-16271.
We gratefully acknowledge useful discussions with V. I. Torgashev.
\end{acknowledgments}

\end{document}